\documentclass{PoS}
\usepackage{mathtools} 
\usepackage{comment}

\title{A Condition Monitoring Concept Studied at the MST Prototype for the Cherenkov Telescope Array}

\ShortTitle{Condition Monitoring at the MST Prototype for CTA}

\author{\speaker{Victor Barbosa Martins},$^a$ Markus Garczarczyk,$^{a}$ Gerrit Spengler,$^b$ and Ullrich Schwanke$^b$ for the MST-STR project of the CTA consortium\\
\llap{$^a$}Deutsches Elektronen-Synchrotron (DESY)\\
Platanenallee 6, D-15738 Zeuthen, Germany\\
\llap{$^b$}Humboldt-Universität zu Berlin\\
Newtonstr. 15, D-12489 Berlin, Germany\\}

\abstract{The Cherenkov Telescope Array (CTA) is a future ground-based gamma-ray observatory that will provide unprecedented sensitivity and angular resolution for the detection of gamma rays with energies above a few tens of GeV. In comparison to existing instruments (like H.E.S.S., MAGIC, and VERITAS) the sensitivity will be improved by installing two extended arrays of telescopes in the northern and southern hemisphere, respectively. A large number of planned telescopes (>100 in total) motivates the application of predictive maintenance techniques to the individual telescopes. A constant and automatic condition monitoring of the mechanical telescope structure and of the drive system (motors, gears) is considered for this purpose. The condition monitoring system aims at detecting degradations well before critical errors occur; it should help to ensure long-term operation and to reduce the maintenance efforts of the observatory. We present approaches for the condition monitoring of the structure and the drive system of Medium-Sized Telescopes (MSTs), respectively. The overall concept has been developed and tested at the MST prototype for CTA in Berlin. The sensors used, the joint data acquisition system, possible analysis methods (like Operational Modal Analysis, OMA, and Experimental Modal Analysis, EMA) and first performance results are discussed.}

\FullConference{36th International Cosmic Ray Conference -ICRC2019-\\
        July 24th - August 1st, 2019\\
        Madison, WI, U.S.A.}

\begin{document}

\section{Introduction}
The Cherenkov Telescope Array (CTA) will be the next-generation gamma-ray observatory, capable of measuring gamma-rays from astrophysical sources with energies ranging from 20 GeV up to more than 300 TeV. \cite{ctareview} The observatory is composed of Imaging Atmospheric Cherenkov Telescopes (IACTs), which measure the nano-second pulses of Cherenkov radiation emitted by the secondary particles of the gamma-ray shower.\cite{hillasreview2013} This detection technique is already well established by current gamma-ray observatories H.E.S.S., MAGIC, and VERITAS.\cite{astrophysicsreview} CTA will provide a larger energy range and sensitivity in comparison to the current observatories.\cite{astrophysicsreview, detailcta} To cover the whole energy range, a design with different sizes and distances between telescopes was defined for two sites, one in the south (European Southern Observatory (ESO) Paranal site, Chile) and one in the north (Instituto de Astrofisica de Canarias (IAC) Roque de Los Muchachos Observatory site in La Palma, Spain). The design foresees the construction of four large-sized telescopes (LSTs) for the lowest energies, 40 Medium-Sized Telescopes (MSTs) for the energy range from 100 GeV to 10 TeV and 70 small-sized telescopes (SSTs) to cover the highest energies. 

\begin{figure}
\begin{center}
\includegraphics[scale=0.5]{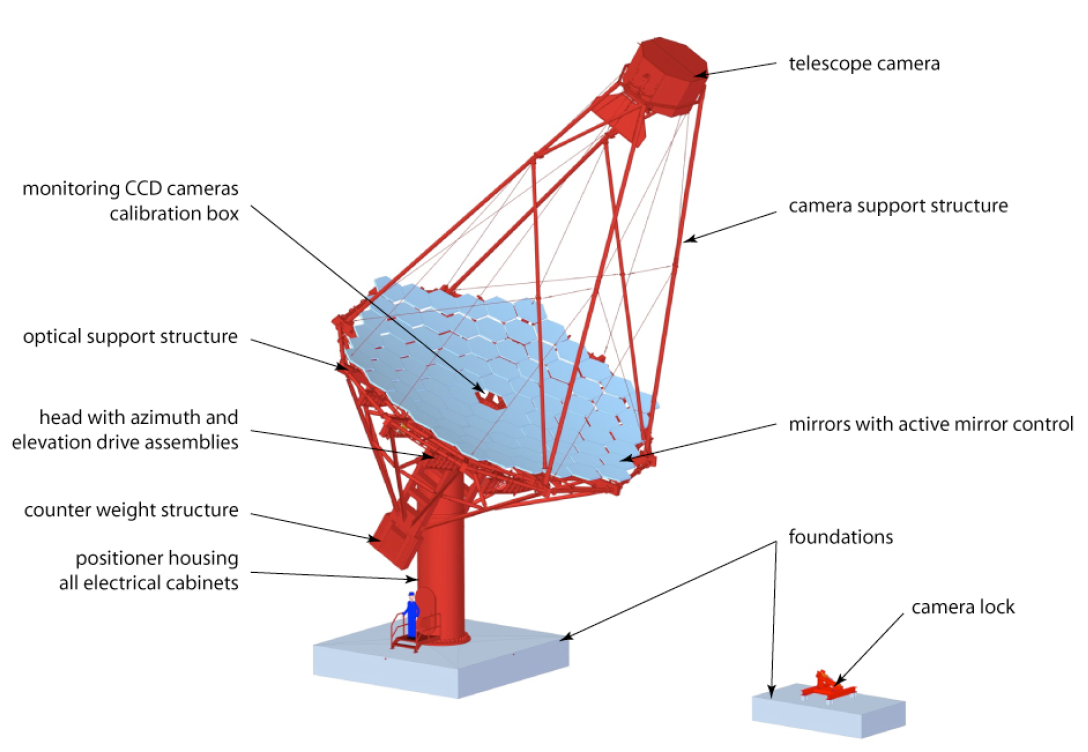}
\caption{Illustrative description of the MST design with the main components.}
\label{fig:mst}
\end{center}
\end{figure}

The MST is based on a modified Davies-Cotton design with a reflector diameter of $12$ m and a focal length of $16$ m as shown in Fig. \ref{fig:mst}.\cite{markusicrc} A prototype was developed, built and is under test in Berlin since 2013. The requirements for the optical, electrical and mechanical components of the MSTs must be fulfilled throughout the 30 years of operation of the observatory. Due to the continuous operation of the telescope, effects such as abrasion, material fatigue, and environmental conditions may affect its performance and reliability. Periodic maintenance in remote sites such as the Roque de Los Muchachos and Paranal is impracticable because of the large number of telescopes and the large area covered by the array. Therefore, a concept for the monitoring of the telescope condition was developed for the MST prototype. 

The Condition Monitoring System (CMS) aims at detecting small changes in the behavior of the telescope, identifying trends in the monitored data and automatically warning the local crew when actions should be taken to avoid the worsening of the telescope performance and prevent major failures to occur. The CMS is divided into Structure Health Monitoring (SHM), detailed in Section \ref{sec:structure} and drive system monitoring (DMS), detailed in Section \ref{sec:drive}. The acquisition system is centralized for both systems and is, therefore, described in Section \ref{sec:daq}.

\section{Structure Health Monitoring (SHM)}
\label{sec:structure}
The goal of the SHM is to detect any change in the structure of the telescope. Every structure which is excited by an external force corresponds to a resonating system and can be, therefore, characterized by its modal frequencies, mode shapes, and damping. A change in the geometry, material or stiffness leads, consequently, to a change in these three features. The overall vibration pattern of the structure is a composition of its individual mode shapes, modal frequencies, and damping. Two methods are very well known and applied nowadays in the industry to derive the modal information from accelerometers datasets: the Experimental Modal Analysis (EMA) and the Operation Modal Analysis (OMA).

The prerequisite to use the EMA is to know exactly the spectrum of the input force acting on the structure and the output force. The frequency response function (FRF) is then calculated by dividing the measured function by the input function. This method is used, for example, in the automotive and aeronautic industry for validation of the manufactured mechanical structures. The input force is usually provided by electronic hammers or electronic shakers. During the experiment, the structure must be first isolated, for example by hanging it, then excited, and measured in many different positions to assure that all the modes are excited and detected.

When the input force is unknown, the OMA must be applied.\cite{palleoma} In this method, the input force is assumed to be a white noise i.e. a broad spectrum excitation. The only way to assure that the method works is to assure: 1) the input spectrum is broad, 2) the force must be applied all over the structure, and 3) the sensors must also be distributed throughout the whole structure. The OMA is applied for large structures, such as bridges, buildings and now telescopes, where it becomes impossible to isolate the structure from external forces, for example, seismic vibration and wind excitation.

In addition to the number and position of the sensors on the structure, the sampling ratio and total acquisition time must be defined according to the frequency range of interest:

\begin{subequations}
\label{eq:thumbs}
\begin{equation}
\centering
f_{s} \geq 2.5 f_{max},
\end{equation}
\begin{equation}
\centering
t_{total} = 1000/f_{min},
\end{equation}
\end{subequations}

\noindent where $f_{s}$ is the sampling frequency, $f_{max}$ is the maximum frequency of interest, $t_{total}$ is the total acquisition time for one dataset, $f_{min}$ is the minimum frequency of interest.

From Finite Element Method (FEM) simulation, it was found that the first modal frequencies of the telescope are between 0 and 10 Hz. According to this frequency range and Eq. \ref{eq:thumbs}, the minimum sampling ratio $f_{s}$ was defined as $100$ Hz and the total acquisition time about $30$ minutes. The data must be acquired with the telescope standing still to avoid any narrowband excitation from the motors of the drive system, which would contradict the OMA assumptions. Besides, the telescope is actually a different structure with different modal parameters at every azimuth and elevation angle configuration. Therefore, a specific configuration (azimuth and elevation equal to zero) was defined to be set up before data acquisition.

\begin{figure}
\begin{center}
\includegraphics[scale=0.65]{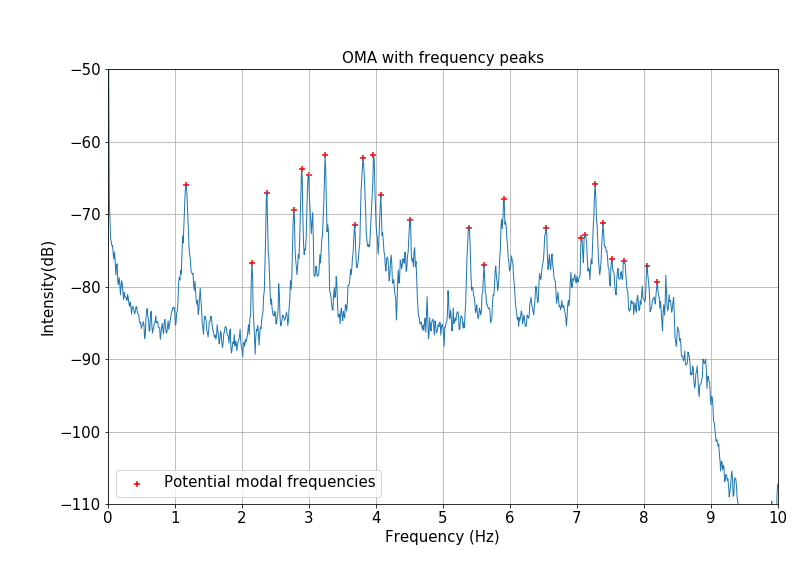}
\caption{Singular value results from a test of the OMA method using only one sensor on the structure. The peaks indicate potential modal frequencies selected by a peak detection technique.}
\label{fig:omagraph}
\end{center}
\end{figure}

After the data acquisition, the OMA is based on the following procedure:\cite{palleoma} detrending, Cross Spectral Density (CSD) calculation, decimation, and Singular Value Decomposition (SVD). During the decimation, every trend on the time series data is eliminated and the mean set to zero. Next, the CSD is calculated between every two measuring channels. Every three channels correspond to the three cartesian coordinates of one sensor. During the decimation, the sampling frequency is decimated from the Nyquist frequency $f_{Ny}=f_{s}/2=50$ Hz to the frequency range of interest of $10$ Hz. This procedure (oversampling + decimation) helps to reduce the noise level in the time series. The SVD is an algebraic decomposition applicable to every matrix to extract the eigenfrequencies and eigenmodes of the system. The CSD matrix for each frequency bin is decomposed in three matrices according to Eq. \ref{eq:svd}:

\begin{equation}
\label{eq:svd}
\centering
CSD_{f} = U.S.V,
\end{equation}

\noindent where U is a matrix $n$x$n$, n being the number of channels, which contains the information about the mode shapes; S is a $n$x$n$ diagonal matrix, which contains the singular values of the system in decreasing order; and $V$ is usually $U^{*}$. If the analyzed frequency is a modal frequency, the first value in the diagonal matrix $s_{0}$ will be much larger than the next values. In this case, the frequency is a modal frequency and the corresponding modal shape is the first column of the U matrix. Figure \ref{fig:omagraph} shows the result of a test made with only one 1-axis force balance sensor (see Section \ref{sec:daq}) placed on the camera frame. Since there is only one measuring channel in the test, the CSD, U, S and V matrixes are all unidimensional. The analysis code was developed by the author in Python and successfully compared to the result from the commercial software Artemis Modal (Svibs). \cite{svibs}

If more sensors were used there would be more curves for the higher degrees of freedom i.e. other singular values in Figure \ref{fig:omagraph} (for details see \cite{palleoma}). A peak detection technique is then applied to the first singular values to extract the potential modal frequencies of the system. After that, a validation method is applied to the selected peaks, the Modal Assurance Criteria (MAC). The method evaluates for each pair of peaks how linearly independent their mode shapes are. If the MAC value for two peaks is higher than a defined threshold (usually >0.85) the peaks correspond to the same mode shape, therefore one of them must be discarded from the selection. The maximum number of linearly independent peaks is equal to the number of channels used in the measurement (degrees of freedom). In this test, for one channel, we could estimate only one modal frequency and mode shape, because all the other peaks would be linearly dependent on the first one.

The next step is to estimate the damping ratio for each modal frequency. For each peak, a MAC value is calculated between the peak frequency and the frequency bins nearby it. Whenever the MAC value is larger than the threshold, the frequency bins correspond to the same mode shape. All the other frequencies should be set to zero. The resulting bell shape curve is then transformed back to the time domain by an Inverse Fourier Transform and the damping is estimated from the decay curve. \cite{damping}

The monitoring system must be basically based on these three features: modal frequencies, mode shapes, and damping ratio. There are different possibilities to monitor them in time: variation in the modal frequencies, or in the difference between two modal frequencies, and variation in the damping ratio. It is also possible to estimate and monitor the distance variation between two sensors, to investigate the Operation Deflection Shape (ODS), which shows how the structure vibrates in each modal frequency, and to use methods for damage detection.

Many of these analysis methods are offered by commercial softwares such as Artemis Modal from Svibs.\cite{svibs} Despite these available solutions in the market, none of them offers the automatization needed for the constant monitoring of 40 telescopes and an effective warning system.  

\section{Drive Monitoring System (DMS)}
\label{sec:drive}
The monitoring of the drive system is a complementary approach to study the status of the telescope. While the SHM studies the telescope when it is standing still, the DMS studies the telescope when it is moving. Any rotating machine vibrates in specific frequencies, which depends on the geometry of the motor and gears and on the rotation speed. The vibration is transferred to the housing and can be also measured by accelerometers. The time series data from the accelerometers is analyzed and the excitation frequencies of the motors can be identified. 

The goal of the DMS is to monitor these frequencies in time to identify trends and changes of the identified frequencies and of the noise level. Severe damages such as total failure of a motor could end up making the telescope unable to park in after a night shift, possibly pointing to the sun, burning the camera and permanently disabling the telescope. It is expected that severe failures come as a consequence of small failures. Damages such as wear, free play, and broken drive teeth could be detected through this method. Since all types of damage in rotating machines result in impact impulses, an increase of the tail of distribution in the spectrum diagram would also indicate an increase in the number of impacts. An automatic warning system will be developed to inform the local crew when some change on the noise level or on the identified frequencies exceeds a threshold level. After that, a maintenance service would be planned to check the status of the motors and they could ultimately be replaced by new motors if needed.

The number of sensors and their positions play an important role in the quality of the data. The sensors must be placed in regions where the vibration is larger, which does not always mean directly on the motor. Experimental tests on the prototype are ongoing. A reproducible procedure for data taking was defined to cover the whole range of possible azimuth and elevation angles and optimized such that the telescope reaches maximum speed at some point during the run. 

Automatic data taking and offline analysis are already running at the prototype telescope in Berlin. Studies on the sensitivity of the sensors are ongoing. Other approaches to investigate the status of the drive are also being explored such as the monitoring of the current, torque and temperature in the motors.

\section{Data acquisition}
\label{sec:daq}
The health of the telescope structure is monitored through accelerometers distributed throughout the Camera Support Structure (CSS), which measure its vibration. Different designs for the numbers, distribution and types of sensors were tested during the last years at the prototype. The force balance sensors were found to be the best choice for the SHM due to their high sensitivity and dynamic range. They are able to measure up to $ng$ vibrations ($1 g = 10 m/s^2$) although the frequency range is limited to some hundreds Hz. This limitation is not prejudicial for the measurement of large structures, once in these cases the modal frequencies are usually of the order of some Hz or even in the sub-hertz regime. These sensors are vastly used in the industry, for instance in the monitoring of bridges, buildings and also the control of seismic vibration.

\begin{figure}
\begin{center}
\includegraphics[scale=0.5]{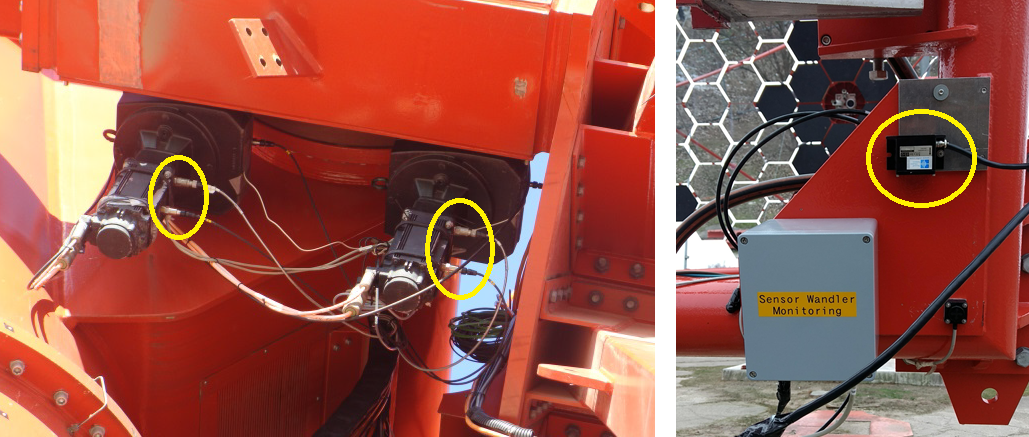}
\caption{Sensors for the condition monitoring system: four accelerometers on the left for the elevation drive monitoring and one sensor on the right for the structure monitoring on the camera frame.}
\label{fig:sensors}
\end{center}
\end{figure}

On the other hand, the sensors used in the DMS must not be as sensitive as the SHM ones but they must be able to sample at kHz rates to monitor the rotation of the motors. Piezoelectric accelerometers usually fulfill this requirement and are used in the prototype for tests: two accelerometers in each of the two azimuth drives (motor and gearbox), and four in each of the two elevation drives. The measuring range of such sensors are of hundreds of g and the frequency range reaches tens of kHz. Figure \ref{fig:sensors} shows on the left four of the DMS sensors and on the right one SHM test sensor with its voltage/current converter.

\begin{figure}
\begin{center}
\includegraphics[scale=0.4]{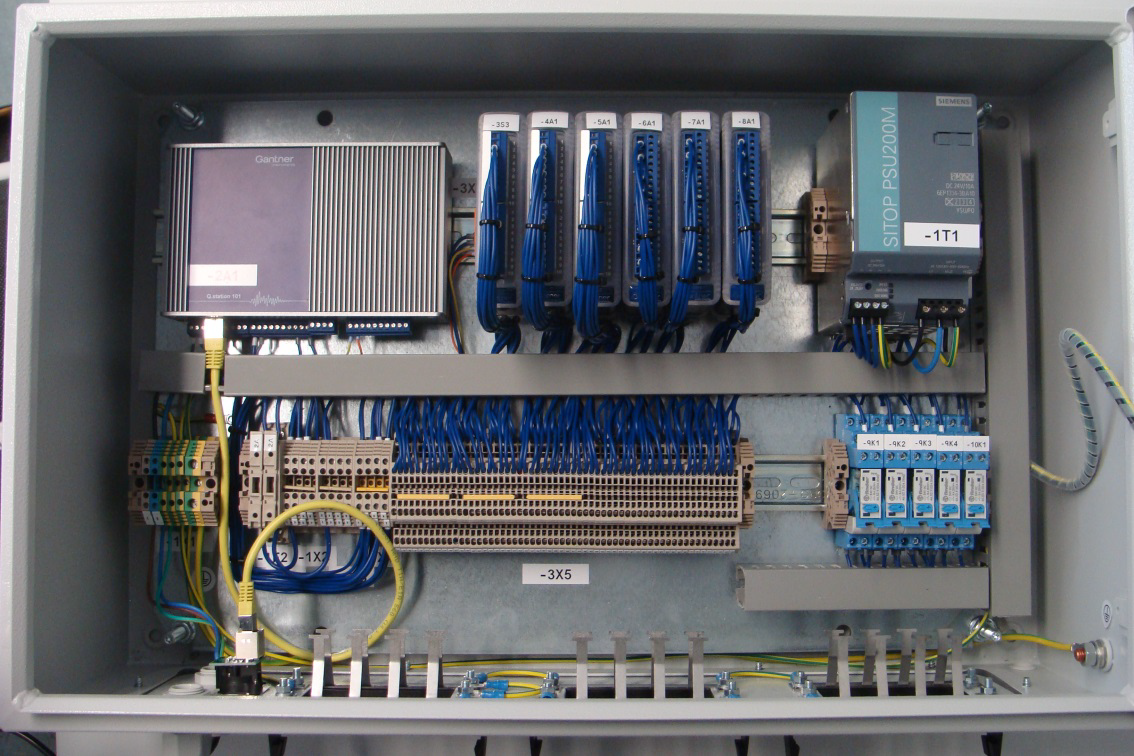}
\caption{Picture of the acquisition system from Gantner Instruments.\cite{gantner} The 6 modules and the Q.station are visible on the upper part of the image from left to right.}
\label{fig:gantner}
\end{center}
\end{figure}

The data acquired by the sensors are converted from analog to digital and processed by modular devices from the company Gantner Instruments.\cite{gantner} The devices are composed of one programmable and self-sustained operating controller Q.station 101, four Q.bloxx A108 modules with 24-bit analog input each, 8 differential channels and 10 kHz sample rate at each channel, and two modules Q.bloxx A111 with four analog inputs each and a sample rate of 100 kHz per channel. The A108 modules are used to process the data from the SHM sensors and the A111 to process the data from the DMS sensors. The controller has an internal hard disk with a storage capacity of one gigabyte for logger functions and offers data transfer via Ethernet. The data is pushed to an FTP server and is then stored in a database for further offline analysis. Part of the analysis runs already automatically in a local machine. Figure \ref{fig:gantner} shows the acquisition system, which is housed inside the telescope tower.

\section{Conclusions and prospects}
\label{sec:conclusions}
The monitoring system for the MSTs is essential to assure that the telescopes are able to fulfill the requirements throughout the whole lifetime of the observatory. By monitoring the structure (SHM) and the drive system (DMS), a complete picture of the status of every MST can be obtained on a daily basis. The analysis is developed based on methods already used in the industry and adapted to our purposes. Besides, the automatic analysis and warning system will spare the need of manpower for the data analysis. Although the results achieved so far are promising, the setup is still experimental and should only reach its final version by the end of 2019. Further development of the analysis code is ongoing and damage tests are also planned for the end of 2019.

\end{document}